\documentclass[doublecol,graphicx,times]{epl2} 
\usepackage{graphicx}

\def\apj{{Ast\-ro\-phys.\ J.}}

\def\grl{{Geo\-phys.\ Res.\ Lett.}}

\def\jgr{{J.\ Geo\-phys. Res.}}

\def\prl{{Phys.\ Rev.\ Lett.}}

\def\pss{{Planet.\ Space Sci.}}

\def\npg{{Nonlin.\ Process.\ Geophys.}}

\def\pop{{Phys.\ Plasmas}}

\def\prl{{Phys. Rev. Lett.}}

\title{Deformation of electron holes in phase space as prerequisite for narrow band electron-cyclotron maser emission}
\shorttitle{Electron phase space hole deformation and radiation} 

\author{R. A. Treumann,\inst{1,2} C. H. Jaroschek\inst{3} \and R. Pottelette\inst{4}}
\shortauthor{R. A. Treumann et al.}

\institute{ 
  \inst{1} Department of Geophysics, Munich University, Theresienstr. 41, D-80333 Munich, Germany\\                   
  \inst{2} Department of Physics and Astronomy, Dartmouth College, Hanover, NH 03755\\
  \inst{3} Department of Earth and Planetary Sciences, Tokyo University, Tokyo, Japan\\
  \inst{4} CETP/CNRS St. Maur des Foss\'es, Cedex, France}

\abstract{
We investigate the role electron holes play in generating fine structure on electron cyclotron maser radiation. It is argued that electron holes become deformed in phase space when interacting with an incomplete ring or horseshoe distribution which occurs in the presence of strong field aligned electric fields in the upward current region and in the presence of a loss cone. This interaction is based on momentum balance considerations. Electron hole deformation causes steep velocity space gradients  that leads to twins of  intense narrow band emission and absorption features that move in frequency space due to the average real space displacement of the deformed electron hole.  
ion.}

\begin{document}

\maketitle
\section{Introduction}
It has been firmly established \cite{ergun2001}  by now that the Auroral Kilometric Radiation (AKR) -- and probably also a number of radio emissions from other celestial objects \cite{ergun2000,begelman2005,treumann2006} -- is due to the (relativistic) electron cyclotron maser in the presence of a magnetic-field-aligned electrostatic potential drop. In its embryonic version \cite{wu1979,melrose1982} the mechanism  referred to loss cone distributions.   However, the potential drop locally evacuates the plasma \cite{chiu1978}
making it manifestly underdense with plasma-to-cyclotron frequency ratio $\omega_{pe}/\omega_{ce}< 1$ and, at the same time, transforms the electron distribution into a ring in velocity space. Realizing this important modification, the electron-cyclotron maser radiation mechanism has been put back on its feet by Pritchett \cite{pritchett1984}. In this realistic version, fundamental emission is generated in the x-mode below but (for not too small frequency ratios) close to the (non-relativistic) cyclotron frequency $\omega_{ce}=eB/m_e$ (where $B$ is the magnetic field strength, $e$ the elementary charge, and $m_e$ the electron mass) and propagates strictly perpendicular to the magnetic field with wave number ${\bf k}\equiv k_\perp{\bf B}/B, k_\|\equiv 0$. 

Controversy  concerns the persistent observation of very narrow-band, very intense emissions of sometimes less than 100 Hz bandwidth, compared to an emission frequency of $\sim 300$\,kHz in the Auroral Kilometric Radiation which corresponds to a relative bandwidth of $\Delta\omega/\omega\sim10^{-4}$. This small relative bandwidth requires extraordinarily steep positive perpendicular phase-space density gradients $\partial f_e/\partial v_\perp>0$ [with $f_e(v_\|,v_\perp)$ the electron phase space density] which in addition must cover a large phase-space volume. The steep gradient is responsible for the  narrow bandwidth, while the large phase-space volume assures high emissivity. Both conditions are difficult to maintain simultaneously and globally. Some models of phase space distributions, meeting these conditions and providing high or marginally high emissivities, have indeed been proposed, for instance by Louarn \cite{louarn1990,louarn1996}, and also in \cite{yoon1998,pritchett2002}. Such models refer to the {\it global} gradients on the electron distribution imposing severe conditions on its shape. Global phase space gradients as sharp as assumed have, to our knowledge, barely been observed. To our opinion the observation of the fast spectral displacement of the emission bands suggests that the radiation sources occupy only a {\it very small} volume in real space. We therefore suggested \cite{pottelette2001,pottelette2005} that the narrow-band spectral fine structure detected in the Auroral Kilometric Radiation is not the result of a steep global gradient in the distribution; rather it is due to emission from `elementary radiation sources'. We tentatively identified these with phase-space (ion and electron) holes of the kind of Bernstein-Green-Kruskal (or BGK) modes investigated in \cite{muschietti1999a,muschietti1999b,singh2000,newman2002,goldman2003},
and others. Justification has been derived from the real-space velocity of the sources deduced from their spectral displacements of the emissions. Even the {\it entire} auroral kilometric emission spectrum may be due to the superposition of the contributions from such tiny `elementary radiators' \cite{pottelette2001}, a proposal that so far is lacking attention. 

Existence of phase-space holes in the downward current region is by now a well established observational fact  that has been backed by numerous numerical simulations \cite{ergun2002,goldman2003}. Observational evidence has also been presented \cite{pottelette2005} for the existence of BGK modes in the low-density upward current  region (along the ambient magnetospheric magnetic field downward directed auroral electron fluxes) where they occur at low frequency manifesting themselves as spikes on the electric wave form.  They had been overlooked so far letting one believe that they would occur only in the downward current region at upward directed auroral electron fluxes along the magnetospheric magnetic field.  

\section{Mechanism of deformation}
\begin{figure}[t]
{\includegraphics[width=7.75cm]{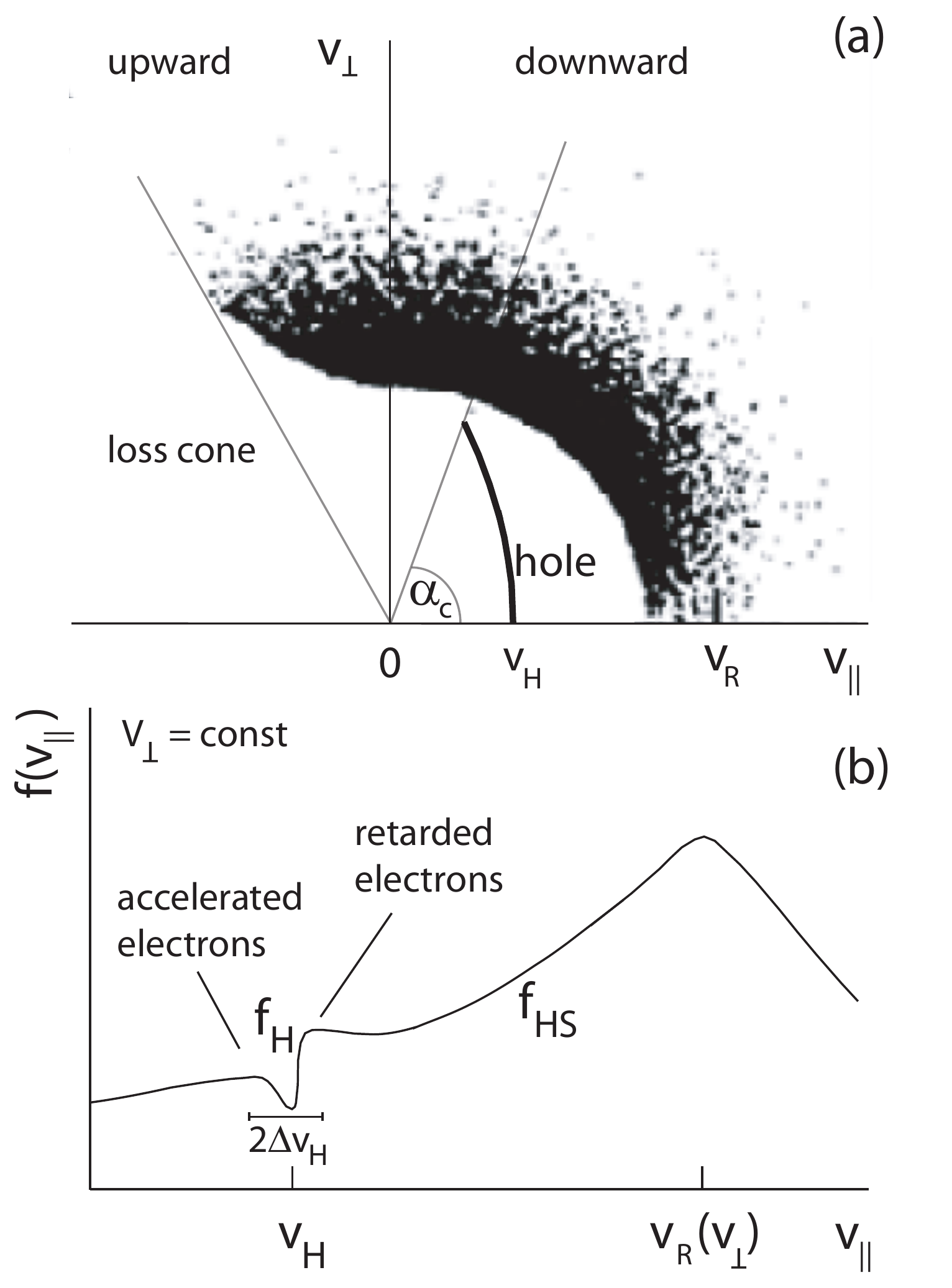}}
\caption{Schematic of the initial shape of an electron hole in phase space. ({\it a}): An example of an incomplete downward hot ring (horseshoe) distribution in phase space with upward loss cone and ring velocity $v_R$ (low-energy particles suppressed). The electron hole (heavy narrow line starting at  velocity $v_H$) has lower velocity $v_H$ and cosinoidal shape which is cut off at the critical angle $\alpha_c$, here assumed to be 70$^\circ$. ({\it b}): Cut through the distribution function $f_{HS}$ at fixed $v_\perp$. The hole causes the dip at $v_H$. On its low-velocity side electrons are accelerated by the hole-electric field; on its high-velocity side electrons are decelerated, causing asymmetric shoulders on the distribution. Since more fast particles are passing, the hole may gain momentum and become accelerated in the direction of high velocities. If the number of trapped particles is constant, the hole deepens and thus grows.}\label{holedef-fig1}
\end{figure}

`Elementary radiators' must be electron holes since ion holes do not directly contribute to radiation even though being important in the dynamics of dilute plasmas \cite{goldman2003}. Being Debye-scale structures, they are beyond resolution of particle detectors and thus invisible on the distribution function. Their phase space extension is determined by their capability of trapping particles. This is restricted to the range $\Delta v_H=\pm\sqrt{2e\phi_0/m_e}$, where $\phi_0$ is the amplitude of the hole potential. Since under real auroral magnetospheric conditions $\phi_0$ is of the order of, say, $\sim 1$ V, the hole is a narrow (and most probably also shallow) distortion on the electron distribution of width $\sim 2\Delta v_H\sim 10^3\,{\rm km/s}$ (Figure \ref{holedef-fig1}b), corresponding to an energy of $\sim 2\,{\rm eV}$ much less than the  $\sim$\,keV downward auroral electron kinetic energies. Any gradients it generates on the distribution function will necessarily be sharp. However, in all models such gradients exist only in the parallel direction. In the following we argue (qualitatively) that electron holes, when suffering deformation in phase space, are capable of providing the required {\it steep perpendicular} phase-space density gradients while occupying modestly large phase-space volumina and in this way directly produce radiation.  

Electron holes deep in the upward current region (i.e. far away from the high density of cold plasma at the edges of the auroral cavity) are excited by a Buneman-like instability \cite{goldman2003} between the (downward flowing) energetic ring (or horseshoe) electrons and the (slow upward) cold ion beam. At a given perpendicular electron velocity $v_\perp= v_{\perp 0}$ = const, the bulk parallel electron velocity of the ring is $v_\|(v_{\perp 0})=v_R\cos\alpha$; $v_R$ is the bulk ring velocity, and $\alpha =\tan^{-1}(v_{\perp 0}/v_\|)$ is the pitch angle between velocity and magnetic field ${\bf B}$. For electrons and ions accelerated by the same parallel electric field the ion beam speed is a factor of $\sqrt{m_e/m_i}$ smaller and can be neglected. The unstable frequency and growth rate of the Buneman instability are $\omega_B\sim (m_e/16m_i)^\frac{1}{3}\omega_{pe}\sim 0.03\,\omega_{pe}$ and $\gamma_B\sim\omega_B$, respectively, and the unstable wave number satisfies the condition $k_B(v_{\perp 0})\sim \omega_{pe}/v_R\cos\alpha$. The initial velocity of the hole at any given angle $\alpha$ is $v_H(\alpha)\approx \omega_B/k_B\sim\,0.03\, v_R\cos\alpha$, much less than $v_R$. For constant $v_R$ it decreases with increasing angle $\alpha$. 

The Buneman instability ceases when the {\it parallel} electron drift velocity approaches the electron thermal velocity. For a ring (or horseshoe) distribution of given thermal spread $v_e$ this occurs at some critical angle $\alpha_c$ above that the instability switches off and the ring excites ion acoustic waves. Thus the shape of an electron hole in velocity space is cosinoidal for $\alpha<\alpha_c$ as shown in Figure \ref{holedef-fig1}a. At larger angles the hole dissolves into trains of ion acoustic waves propagating parallel to the magnetic field at the corresponding ion-sound speeds. Because of this cosinoidal shape an electron hole in phase-space is in fact a two-dimensional bent entity, itself exhibiting steep gradients in both directions, parallel and perpendicular to the magnetic field. 
\begin{figure} [t]
{\includegraphics[width=7.75cm]{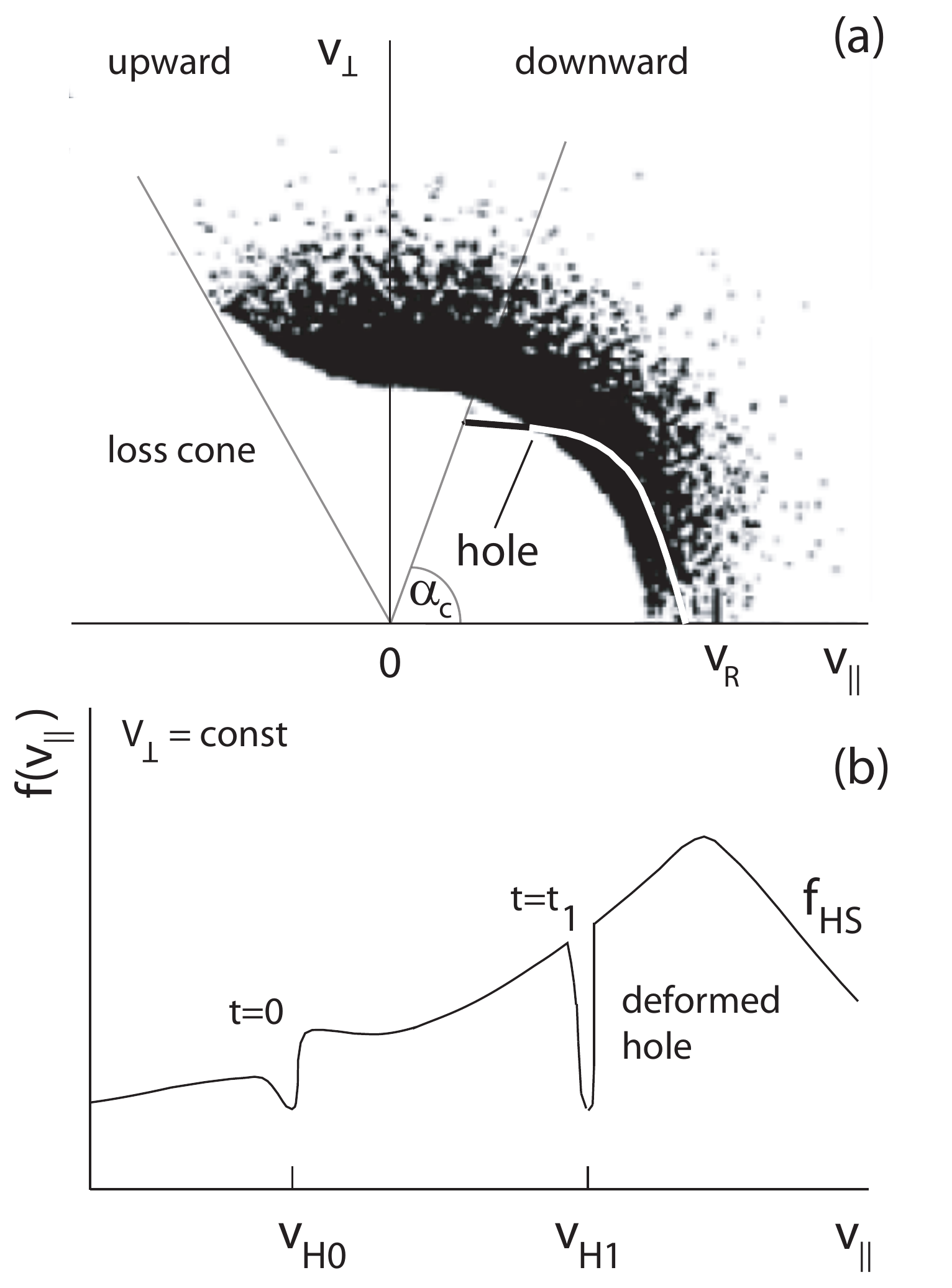}} \caption{Example of hole deformation. 
({\it  a}):
Shape of the deformed hole in velocity space. The hole has been shifted into the bulk of the distribution at different rates for each perpendicular velocity. Shift and deformation produce steep perpendicular velocity space gradients at the hole edges both in parallel and perpendicular directions. ({\it b}): Distribution at constant $v_\perp$ showing the position of the hole at time $t=0$ and at a later time $t=t_1$ when the hole moves into the bulk of the distribution. Steep gradients evolve for constant hole-phase-space density.}\label{hhole-fig2}
\end{figure}

These perpendicular gradients are the important ingredient in the electron-cyclotron maser radiation mechanism. They do already contribute to radiation. However, being located on the low-phase-space density low-velocity side of the electron distribution, the phase-space gradients are still moderate, the holes are shallow, and the phase-space volume is too small for them to provide high emissivities. 

For this to achieve one wishes the hole to be displaced deep into the bulk of the distribution function. This requires some kind of parallel-momentum exchange between the distribution and the hole for which no mechanism has been identified yet. Would the hole be fixed in space and of symmetric (positive) electric potential  $\phi(z)$, the net momentum exchange between the hole and the passing electrons would vanish, for the deceleration and acceleration of any passing electron cancel exactly. However the hole floats in space and by itself reacts to the passing electrons. When an electron enters the hole on one side it tends to decelerate the hole; when, on the other hand, it leaves the hole, it tends to drag the hole into its own direction of motion. This causes the hole to oscillate. Because being massive and of density $n_H$, the hole is inert. Its reaction time is given by the inverse of its proper plasma frequency $\tau_H\sim\omega_{pH}^{-1}$, where $\omega_{pH}^2=e^2n_H/\epsilon_0m_e$ which is small since $n_H\ll n_e$. The reaction time is thus quite long. Fast electrons with transition time $\tau_e=2\Delta/v\ll\tau_H$ do not take notice of the hole, and the hole, in the average, does not feel them before they escape. 

In order to have an effective momentum exchange the electron distribution must be asymmetric at the location of the hole. When the phase-space distribution increases towards positive velocities, i.e. when $[\partial f_{hs}/\partial v_\|]_{v_\|=v_H} >0$, then this has the effect that on its trailing (upward) edge (in the hole frame) the hole feels the drag of only few  fast upward electrons. In contrast, on its leading (downward) edge it is subject to the drag of the many fast downward electrons. Both groups of particles tend to accelerate the hole by attraction, but since there are more electrons moving downward than upward, the hole is pulled downward until it approximately catches up with the bulk velocity of the electrons. 

By moving toward the high speed part of the distribution the hole maintains its density. The dip it has bitten out of the distribution deepens thereby, the hole amplitude increases, and the phase-space density gradients steepen. This happens at each fixed perpendicular velocity with the effect decreasing with increasing $v_\perp$ and at $\alpha_c$ vanishing at all. Hence the hole becomes increasingly stronger deformed at smaller angles $\alpha<\alpha_c$. This deformation and the related steepening of the phase space gradient is sketched in Figure \ref{hhole-fig2} where it is shown for the parallel gradient  $\partial f_e/\partial v_\|$ in the lower part of the figure. Because at $\alpha_c$ the hole is at rest, at angles $\alpha$ closer to $\alpha_c$ the parallel gradient transforms completely into a perpendicular gradient $\partial f_e/\partial v_\|\to \partial f_e/\partial v_\perp$. Thus it is the deformation of the hole that generates the steep velocity space gradient on the electron distribution function, and this gradient exists only at the location of the electron hole in real space.

\section{The effect on the radiation} 
For the generation of radiation it is important that the deformation of the hole in phase-space causes a curvature of the hole shape that leads to perpendicular velocity gradients $\partial f_H/\partial v_\perp$ as steep as the parallel velocity gradients. However, $\partial f_H/\partial v_\perp>0$ is found only on the high-speed edge of the hole; on the low-velocity edge $\partial f_H/\partial v_\perp <0$. While the high-velocity side of the hole generates radiation, radiation will be reabsorbed at the low-velocity side of the hole. 

Because of the small velocity spread of the hole, emission and absorption are not at the same frequency, however. The hole-velocity spread is $\Delta v_H\equiv |v-v_H|\simeq 2(2e\phi_0/m_e)^\frac{1}{2}$, where $\phi_0$ is the maximum of the electrostatic hole potential. The (relativistic) resonance condition for strictly perpendicular emission (i.e. the resonance circle)
\begin{equation}\label{eq:rescond}
 v_\perp^2 + v_\|^2 = c^2(1-\omega^2/\omega_{ce}^2)
\end{equation}    
at fixed parallel velocity $v_\|=$\,const requires that the absorption occurs at frequency $\omega_{ab}>\omega_{em}$ higher than the emission frequency $\omega_{em}$ by the small amount
\begin{equation}\label{eq:deltaomega}
\frac{\Delta \omega}{\omega_{em}}\simeq \frac{4 v_\perp}{c}\frac{\omega_{ce}^2}{\omega_{em}^2}\sqrt{\frac{2e\phi_0}{m_ec^2}}
\end{equation}
Since $\omega_{em}\simeq\omega_{ce}$, $\phi_0\simeq$\, 1\,V, and $v_\perp/c\simeq 10^{-2}-10^{-1}$, the absorption occurs at a few Hz up to a few 10 Hz above the emission. This coincides with the occasional claim that the Auroral Kilometric Radiation emission has bandwidth down to 10 Hz. Current instrumental resolution is marginally capable of detecting such narrow structures. Unambiguous detection of twins of emission/absorption patterns would provide a strong indication of emission from deformed electron holes.  

\begin{figure}[t]
\hspace{-0.4cm}{\includegraphics[width=7.75cm]{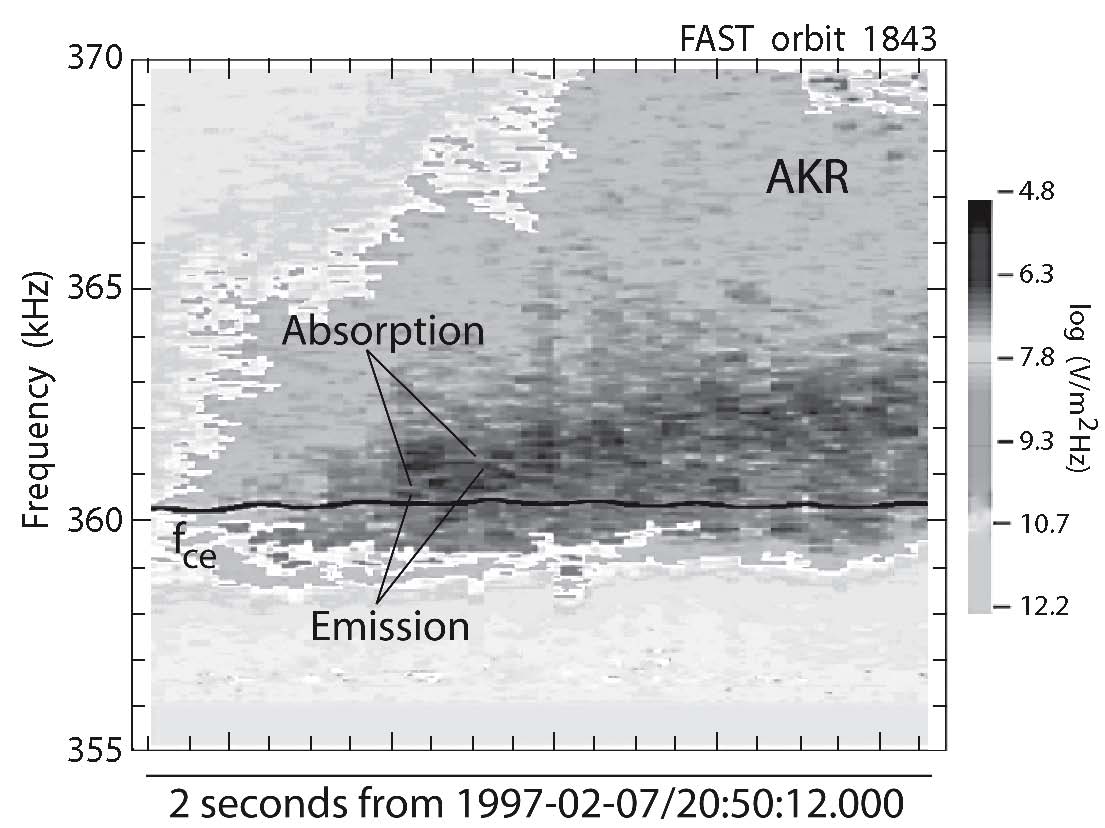}}
\caption{Two seconds of a very high time-resolution dynamical source region spectrum
of AKR. The emission is partially local as indicated by the
frequency being below the electron cyclotron frequency (black
line). The fine structure of AKR suggests that it consists of many
small-scale elementary radiators causing extremely narrow-band emissions and absorptions. Two such emission/absorption-twin features have been indicated.}\label{bhole-fig3}
\end{figure}

Figure \ref{bhole-fig3} shows an example of a highest time-resolution spectrum of FAST observations when the spacecraft passed  the Auroral Kilometric Radiation source region \cite{pottelette2005}. This spectrum indeed exhibits groups of adjacent emission/absorption features of the kind expected for electron-hole radiation. The bandwidth of some of the groups is less than 100 Hz,  in agreement with the predictions. Here in the Auroral Kilometric Radiation source region the groups seem to be short-lived suggesting that the emission process is highly dynamical. 

Emission and absorption features are in fact seen both below and above the local cyclotron frequency. Radiation below $\omega_{ce}$ will preferentially be strictly perpendicular. While radiation at slightly higher frequency could originate from slightly beneath spacecraft, the `elementary radiator' mechanism offers another explanation that can also be based on the deformation model.

The resonance line Eq. (\ref{eq:rescond}) of the electron-cyclotron maser for strictly perpendicular emission below the electron cyclotron frequency is a perfect circle. Deformed electron holes do not exhibit circular shape in phase space. They do not map one single resonance circle. They can rather be approximated piecewise by a set of resonance circle elements of which the best fitting part contributes to highest emissivity.  On the other hand they can also be approximated by sections of shifted ellipses in which case they generate {\it oblique} radiation of frequency close to but not necessarily below the electron cyclotron frequency; rather the emitted frequency is higher than $\omega_{pe}$ because of the modified resonance condition which now includes a finite parallel component of the wave number $k_\|\neq 0$. This might be the case for the observed narrow-band emission/absorption features just above $\omega_{ce}$ in Figure \ref{bhole-fig3} as well as in other places. 

In the more general case, on the other hand, it will not be possible to map strongly deformed holes to one single resonance ellipse nor resonance circle. In this case their edge regions might piecewise contribute to radiation and absorption at different frequencies. Nevertheless, because of the very steep velocity space gradients involved at the different edge pieces (see Figure \ref{hhole-fig2}), the emissivity at each of the frequencies will not be small even though only small phase space volumes become involved in the generation of the radiation. The steepness of the gradient compensates for the lack of a large phase space volume in this case. Oblique radiation of this kind can much easier escape from the large-scale density cavity into free space due to multiple reflection at the cavity boundaries than strictly perpendicular emissions.

All the narrow emission bands observed so far exhibit a well pronounced even though apparently irregular motion across the spectrum. These spectral displacements have traditionally been identified with the real-space motion of the radiation source \cite{pottelette2001}. In the deformation model such a motion could also be caused by the phase-space displacement. In fact, momentum exchange between $f_{HS}$ and $f_H$ is expected to be strongest at small $v_\perp$. Hence, the hole will move initially fastest at low $v_\perp$ readily reaching its final $v_\|$-position near $v_\perp\sim 0$. Only after having settled there the hole shape will deform at higher $v_\perp$ as well. At fixed (final) $v_\|$ therefore $v_\perp$ increases and, from the resonance condition Eq. \ref{eq:rescond} for strictly perpendicular radiation the frequency should decrease. However, since this effect is small in a weakly relativistic plasma, it probably causes only a small displacement of the emitted frequency towards smaller values and will not be distinguished from the stronger displacement of the hole in real space. Therefore, the observed spectral motion of the emission band may still be safely attributed to the real-space displacement of the hole. 

The bulk hole velocity is calculated from the \revision{first} moment of the hole distribution function $f_H$ as 
\begin{equation}\label{holevel}
n_H v_H=\int\limits_{\rm hole} v_\|v_\perp^3{\rm d}v_\|{\rm d}v_\perp f_H(v_\|,v_\perp)
\end{equation}
where the hole distribution is subject to the reduced kinetic equation
\begin{equation}\label{eq:holeeq}
\frac{{\rm d}f_H}{{\rm d}t} = -\frac{e}{m_e}E_\|(z-z_H)\left.\frac{\partial f_{HS}}{\partial v_\|}\right|_{v_\perp }
\end{equation} 
This expression follows from the Vlasov equation and the definition of the electron distribution $f_e=f_{HS}-f_H$. $E_\|$ is the self-consistent parallel electric field centered at the temporal position of the hole, $z_H(t)$, which itself is a function of time. 

This equation must be solved for the dynamics of the passing electrons (to first approximation neglecting the effect of trapped electrons, keeping the density $n_H$ of the hole constant), which follows from $m_e{\dot v} =- eE_\|(z-z_H)$ and $v={\dot z}$ with $\epsilon_0\nabla_\|E_\|=e(n-n_{HS}+ \int {\rm d}v_\| f_H)\simeq e\int {\rm d}v_\| f_H$. 

In order to account for the de-correlation of the hole response, the above mentioned condition on the transition time must be worked out. Thus integration extends only over times $t <\omega_{pH}^{-1}$ for particles not becoming trapped. 
The integral in Eq. (\ref{holevel}) is to be taken over the entire deformed shape of the hole in phase space. Hereby the average motion of the hole can result to be directed in any direction, upward or downward, depending on the shape and contribution of the hole distribution $f_H$ to the integral. 

\section{Conclusion}
The present discussion is intended to support the view that deformation of electron holes in phase space is capable of generating very narrow band fine structure in radiation spectrum that is emitted in the electron-cyclotron maser emission mechanism. Such  fine structure has been observed in the Auroral Kilometric Radiation and poses a serious problem for explanation by conventional theories.
It has often be stressed that for the electron cyclotron maser to work the electron distribution function should exhibit the presence of a large and empty loss cone. The ordinary ring maser works indeed completely without such a loss cone when generating perpendicular radiation beneath the local electron cyclotron frequency in the x-mode. Such a ring distribution is ideally symmetric. As we have seen, symmetric distributions will not have any effect on the deformation of electron holes leaving them undeformed. Consequently, they will also not generated fine structure in the emitted radiation. 
The observation of the narrow band fine structure in the emitted spectrum and the presence of twins of emission and absorption lines suggests that the distribution function must be non-symmetric with respect to positive and negative parallel velocities $v_\|$. This is just the case in the presence of a loss cone. It may thus be concluded that the presence of a loss cone is indeed most important for the electron cyclotron mechanism to work though in a different sense than believed. It is not the loss cone that generates the emitted radiation. However, without the presence of a loss cone the ring distribution would be symmetric and would be unable to affect the structure of the electron holes. With the loss cone present its symmetry is broken, and it become a horseshoe distribution as is observed. It is such distributions which are able to produce the elementary radiators and the observed fine structure. The presence of a loss cone is thus indeed of essential importance for the electron cyclotron maser. 

So far we have not specified the very process of momentum exchange in phase space. This can only be done by numerical simulation and will be the purpose of a follow-up communication on this subject [C.-H. Jaroschek et al., in preparation, 2008]. 
The discussion in the present Letter neglects a large number of effects which happen to take place in hole dynamics: particle acceleration and retardation when interacting with the hole, particle trapping, hole stability and internal dynamics, dissipation of the hole electric field, and the various processes of wave excitation in the hole interior and in hole interaction. We are very well aware of the importance of these effects in the required momentum exchange that is based on the de-correlation we have proposed.  Various authors have investigated one or the other of them, but the important question of hole deformation in phase space has to our knowledge not yet been treated anywhere. 

\acknowledgments This research is part of a Visiting Scientist Programme at ISSI, Bern. CHJ has benefited from a JSPS Fellowship of the Japanese Society for the Promotion of Science. CHJ and RT thank M. Hoshino for his hospitality, support and discussions. This research has also benefitted from a Gay-Lussac-Humboldt award of the French Government which is deeply acknowledged. The data access was through the French-Berkeley cooperative program on FAST supported by the French PNST programme.

\end{document}